\documentclass[pra,showpacs,twocolumn,groupedaddress,amsmath,amssymb]{revtex4}

\usepackage[american]{babel}
\usepackage{graphicx}
\usepackage{times}
\usepackage{color}

\newtheorem{theorem}{Theorem}
\newtheorem{lemma}{Lemma}

\begin{document}
\title{Nonclassicality filters and quasiprobabilities}

\author{T. Kiesel and W. Vogel}

\affiliation{Arbeitsgruppe Quantenoptik, Institut f\"ur Physik, Universit\"at  Rostock, D-18051 Rostock,
Germany}

\begin{abstract}
Necessary and sufficient conditions for the nonclassicality of bosonic quantum states are formulated by introducing nonclassicality filters and nonclassicality quasiprobability distributions. Regular quasiprobabilities are constructed from characteristic functions, which can be directly sampled by balanced homodyne detection. 
Their negativities uncover the nonclassical effects of general quantum states. The method is illustrated by visualizing the nonclassical nature of a squeezed state.
\end{abstract}
\pacs{03.65.Ta, 42.50.Dv, 03.65.Wj}

\maketitle

\section{Introduction}

 The foundations of quantum theory have been known for several decades, but the relation to classical physics is still a topic of current research. In quantum optics, the notion of nonclassicality caused  long-lasting discussions. The quantum state of a radiation field is often examined by means of photodetectors which measure normally ordered field  correlation functions. The latter are properly described by the quasiprobability distribution or $P$ function of Sudarshan and Glauber~\cite{prl-10-277,pr-131-2766}. Following Titulaer and Glauber~\cite{pr-140B-676}, 
''states with positive $P$~functions \dots are \dots possessing classical analogs.'' The other way around, a quantum state is nonclassical if the $P$ function does not exhibit the properties of a classical probability distribution, cf. e.g.~\cite{scr-T12-34}. 

Any quantum state of a harmonic oscillator can be given as a quasimixture of coherent states $|\alpha\rangle$,
\begin{equation}
\label{eq:P-repr}
{\hat \rho}=\int d^2 \alpha P(\alpha) |\alpha\rangle \langle \alpha|,
\end{equation}
where $P(\alpha)$ is the $P$~function mentioned earlier, cf.~\cite{prl-10-277,pr-131-2766}. The coherent state is known to be that quantum state which is most closely related to the classical behavior of an oscillator. Its $P$~function is formally equivalent to the deterministic classical phase-space distribution, representing a single point in phase space. If the $P$~function has the properties of a classical probability density, $P(\alpha)\equiv P_{\rm cl} (\alpha)$, the state is a true classical mixture of coherent states.  Hillery has shown that the coherent states are the only pure quantum states having a non-negative $P$~function~\cite{pl111a-409}. Hence, for any classical mixture of coherent states the $P$~function exactly reflects the classical behavior of the oscillator in phase space -- including its free evolution. The failure of the interpretation of $P(\alpha)$ as a probability density, $P(\alpha)\not= P_{\rm cl} (\alpha)$, is  intimately related to the quantum superposition principle, thus it most naturally displays the quantumness of any quantum state.

However, in general the $P$~function can only be understood as a generalized function which is often  not accessible. For this reason, different representations of a quantum state are considered. An often used one is the Wigner function~\cite{pr-40-749}, which also covers the full information on the quantum state. A generalization yields the set of $s$-parameterized quasiprobability distributions~\cite{pr-177-1882}. By fixing the parameter~$s$, different quasiprobabilities are obtained. If one of these functions violates the requirements of a classical probability distribution, the given state is nonclassical. Unfortunately, this set of functions does not reveal all nonclassical effects in terms of regular functions: For a squeezed state, they are either non-negative or highly singular. In order to develop quasiprobabilities
to uncover nonclassicality in general, the generalized quasiprobabilities of Agarwal and Wolf will be a powerful foundation~\cite{prd-2-2161}. 

Another general representation of a quantum state is its characteristic function, defined as the Fourier transform of a given quasiprobability. Its advantage lies in the fact that it is always a regular function, even the characteristic function of an irregular nonclassical $P$~function. Useful nonclassicality conditions have been derived~\cite{prl-84-1849,prl-89-283601}
and applied in experiments~\cite{pra65-033830,pra75-052106,pra-79-022122}.
However, for a full characterization of nonclassicality one needs to check an infinite hierarchy of conditions, which may be a cumbersome procedure. We will use them as the starting point for our examination.

In this article we introduce regular quasiprobabilities with the aim to uncover all types of nonclassical effects by their negativities. A distribution of this type, to be called nonclassicality quasiprobability, belongs to the set of the Agarwal-Wolf quasiprobabilities. For our purposes the filter functions occurring in the latter must obey specific constraints. We study the properties, which are needed to make the
filters useful for experimental applications, and we show how to construct them. 

The article is structured as follows. In Sec.~II we introduce the requirements for general nonclassicality filters and discuss the relation to previously known filter procedures. Section~III is devoted to nonclassicality quasiprobabilities, which, in addition, contain full information about the quantum state. The method is illustrated for the example of a squeezed vacuum state. In Sec.~IV we briefly summarize our results.

\section{Nonclassicality filters}
\subsection{Characteristic functions and Bochner's theorem}

Let us now consider the possibility of getting general insight into the properties of the $P$~function in an experiment. 
The characteristic function $\Phi(\beta)$, defined as the Fourier transform of 
$P(\alpha)$, can be sampled by balanced homodyne detection, cf.~\cite{pra-78-021804,pra-79-022122}. From a set of quadrature data $\{x_j[\varphi]\}_{j=1}^N$ at some fixed phases $\varphi$, it can be estimated by
\begin{equation}
\label{eq:sampling}
        \Phi(\beta) = \frac{1}{N} \sum_{j=1}^N e^{i |\beta| x_j[\pi/2-\arg(\beta)]} e^{|\beta|^2/2}.
\end{equation}
Thus we have direct experimental access to the characteristic function $\Phi$.
It may be a rising function of $|\beta|$, whose Fourier transform only exists as a highly singular distribution~\cite{pra-79-022122}. The standard deviation of $\Phi$ is, for a given sample of data, bounded by~\cite{pra-78-021804}
\begin{equation}
        \sigma\{\Phi(\beta)\} \le \frac{e^{|\beta|^2/2}}{\sqrt{N}}. 
	\label{eq:std:dev}
\end{equation}
Interestingly, also, the characteristic function $\Phi(\beta)$ of any quantum state is bounded by $e^{|\beta|^2/2}$~\cite{Perelomov}.

If $\Phi(\beta)$ is not square integrable, its Fourier transform is highly singular. In such cases, the nonclassicality of the quantum state can be identified via Bochner's theorem~\cite{math-ann-108-378}. 
The function $\Phi(\beta)$ is in general continuous, with $\Phi(0) = 1$, $\Phi(-\beta) = \Phi^*(\beta)$. The $P$ function has the properties of a probability density if and only if for all positive integers $N$ and complex $\beta_1,\ldots,\beta_N$, the matrix $\left(\Phi(\beta_i-\beta_j)\right)_{i,j=1,\ldots,N}$ is positive semidefinite. This leads to an infinite hierarchy of nonclassicality conditions~\cite{prl-89-283601}, which in practice cannot be examined completely. However, for $N=2$ we obtain a simple inequality, which is valid for all $\beta$ and necessary for classicality~\cite{prl-84-1849}: 
\begin{equation}
|\Phi(\beta)| \leq 1. \label{eq:Bochner:2}
\end{equation}
The violation of this inequality can be used to experimentally demonstrate the nonclassicality of a quantum state \cite{pra65-033830,pra75-052106,pra-79-022122}. 

In cases when the inequality~(\ref{eq:Bochner:2}) is fulfilled, one cannot directly infer classicality, but there is a chance that the characteristic function is square integrable. A prominent example are the photon-added thermal states~\cite{pra-46-485}. Then one can perform the Fourier transform to obtain the $P$~function and check nonclassicality by its original definition~\cite{pra-78-021804}. Severe problems occur to identify nonclassicality 
if the characteristic function satisfies~(\ref{eq:Bochner:2}), but is not square integrable.

\subsection{Filtered $P$ functions}  

Let us now develop a simple and general method for identifying the nonclassicality of a quantum state under realistic experimental conditions.
It is based on filtering of the characteristic function,
\begin{equation}
\label{eq:Phi-filter}
        \Phi_\Omega(\beta;w) = \Phi(\beta) \Omega_w(\beta),
\end{equation}
with a filter function $\Omega_w(\beta)$, which we will allow to depend on a real parameter $w$. The filter shall satisfy the following specific properties:
\begin{enumerate}
		\item[(a)] {\em Universality.} For any quantum state, the filtered characteristic function $\Phi_\Omega(\beta;w)$ is square integrable, such that its Fourier transform, $P_\Omega(\beta;w)$, is a well-behaved function. Since $\Phi(\beta)$ and its standard deviation (cf.~Eq.~(\ref{eq:std:dev})) are bounded by $e^{|\beta|^2/2}$,  we need that $\Omega_w(\beta)e^{|\beta|^2/2}$ is square integrable for all $w$. This ensures that the method is universal: it applies to any quantum state and to realistic  experimental data.
        \label{en:cond:universal}
        \item[(b)] {\em Non-negativity.} To detect nonclassicality of unknown quantum states by negativities in the regularized function $P_\Omega(\beta;w)$, the latter shall be non-negative for all classical states. Equivalently, the filter $\Omega_w(\beta)$ shall not cause additional negativities in the regularized function $P_\Omega$. This requires that $\Omega_w(\beta)$ itself has a non-negative Fourier transform.
\label{en:cond:non-negative}
        \item[(c)] {\em Completeness with respect to the nonclassicality of the $P$~function.} The parameter $w$ represents the width of the filter. It may be introduced by a scaling transform,
        \begin{equation}
                \Omega_w(\beta) = \Omega_1(\beta/w).
        \end{equation}
For an infinitely wide filter, the $P_\Omega$~function approaches the original $P$~function. This requires, for all $\beta$, that 
        \begin{equation}
\lim_{w\to\infty}\Phi_\Omega(\beta;w) = \Phi(\beta),
        \end{equation}
or equivalently, 
$\Omega_w(0) = 1$ and $\lim_{w\to\infty} \Omega_w(\beta) = 1$.
        \label{en:cond:limit}
\end{enumerate}

The most simple example of such a filter is a two-dimensional triangular filter, $\Omega_w(\beta_r+i\beta_i) = {\rm tri}(\beta_r/w) {\rm tri}(\beta_i/w)$, where ${\rm tri}(x) = 1 - |x|$ for $|x| < 1$ and ${\rm tri}(x) = 0$ elsewhere. 
Since this function has compact support for all $w > 0$, it satisfies the condition~(a). Furthermore, it obeys the constraints~(b) and~(c), since the Fourier transform of the triangular function is non-negative and $\lim_{w\to\infty}\Omega_w(\beta) = 1$, respectively.

This example clearly shows that there exist filters which satisfy all our requirements. Most interestingly, they can be used to detect nonclassicality of any nonclassical state. The other way around, the negativities are uniquely caused by the nonclassicality of the state, not by the filter.  For all nonclassical states, we can find a regularized function $P_\Omega$ which displays negativities. We refer to such filters $\Omega_w(\beta)$ as nonclassicality filters.  For the proof of their general properties, we refer the readers to theorem~1 in Appendix~\ref{app:A}.

\subsection{Relation to known filtering procedures} 
Filtering procedures of the $P$~function having the structure of Eq.~(\ref{eq:Phi-filter}) are already known. However, there is no procedure known that fulfills all the requirements~(a)--(c).
Let us briefly consider such filtering approaches together with their shortcomings for nonclassicality detection. Note that the following approaches had not been designed for that purpose.

By choosing $\Omega_s(\beta) = \exp((s-1)|\beta|^2/2)$ as a family of filters, we consider the $s$-parameterized quasiprobabilities~\cite{pr-177-1882}. For $s = 0$, we get the Wigner function; for $s = 1$, the $P$~function; and for $s=-1$, the $Q$~function. It is obvious that such filters do not fulfill the universality condition~(a) for $s \geq 0$. Therefore, they are not capable of regularizing the $P$ function of an arbitrary state for $s > 0$. There exist nonclassical states which do not possess a regular $s$-parameterized quasiprobability showing negativities. Squeezed states are a prominent example. If their nonclassical effects would already be displayed for $s = 0$, they could be observed as negativities in the Wigner function. However, the Wigner function of a squeezed state is always non-negative.

Another filter was considered by Klauder~\cite{prl-16-534}. He showed that appropriate filtering of the $P$~function may lead to an infinitely differentiable regular function. This filtering was recently applied to regularize the $P$~function of a squeezed state~\cite{science-322-563}. However, since Klauder's  filtering does not obey the non-negativity condition~(b), the corresponding negativities of the regularized functions are not uniquely related to the nonclassicality of the considered quantum state.

Last but not least, a very general approach to define quasiprobabilities
and operator ordering was introduced by Agarwal and Wolf~\cite{prd-2-2161}. This may be considered as a general background of our considerations. Since the authors' aim was to provide general methods, they did not 
consider constraints of the type needed for the nonclassicality filtering.

\section{Nonclassicality quasiprobabilities}
\subsection{Filters for quasiprobabilities}

Filters with compact support can be easily applied to experimental data. However, one loses information about the quantum state, such that the latter cannot be recovered completely from a filtered $P$ function. To overcome this problem, one has to use invertible nonclassicality filters. They have to meet the criteria of Agarwal and Wolf~\cite{prd-2-2161}, in particular, having no zeros anywhere, in order to preserve all information about the state. 

To our knowledge, no simple examples for such filters are known, but they can be constructed in the following way:  Let us assume that some positive continuous function $\omega(\beta)$ satisfies $\omega(-\beta) = \omega(\beta)$ and decays sufficiently fast; that is, $\omega(\beta) e^{u|\beta|^2}$ is square integrable for any $u > 0$. For example, one may choose
\begin{equation}
\label{eq:ac-ex}
\omega(\beta)=\exp(-|\beta|^4).
\end{equation}
It is easy to see that its autocorrelation function, 
\begin{equation}
\label{eq:ac}
        \Omega(\beta) =\tfrac{1}{\mathcal N}\int \omega(\beta')\omega(\beta+\beta')\, d^2\beta', 
\end{equation}
with $\mathcal{N} = \int |\omega(\beta)|^2\, d^2\beta$, is positive and satisfies $\Omega(-\beta) = \Omega(\beta)$.  Moreover, we find that $\Omega(0) = 1$ and the Fourier transform of an autocorrelation function is always non-negative. Finally, it decays sufficiently fast, so that~$\Omega(\beta) e^{u|\beta|^2}$ is square integrable for any $u > 0$(cf.~lemma~1 in Appendix \ref{app:B}). 

Now we define a set of functions by
\begin{equation}
\label{eq:acw}
        \Omega_w(\beta) = \Omega(\beta / w), \qquad w > 0.
\end{equation}
Since $\Omega_w(\beta)$ is continuous, the sequence of functions converges for all $\beta$ pointwise to $1$ when $w\to\infty$:
\begin{equation}
\lim_{w\to\infty}  \Omega_w(\beta) = \lim_{w\to\infty}\Omega(\beta/w) = \Omega(0) = 1.
\end{equation}
Hence these functions satisfy all criteria for being a nonclassicality filter. Since $\Omega_w(\beta)$ has no zeros, the regularized function $P_\Omega$ contains all the information about the quantum state, and consequently, it represents a generalized phase-space function in the sense of Agarwal and Wolf~\cite{prd-2-2161}. Therefore, for any nonclassical state, one can find a regular quasiprobability distribution which displays the nonclassical character by its negativities. We refer to such distributions as nonclassicality quasiprobabilities.


The experimental implementation of the procedure to identify nonclassicality of a general and unknown quantum state is straightforward:
\begin{enumerate}
        \item[(A)] {\em Sampling.} Direct sampling of the function $\Phi(\beta)$ from experimental data, Eq.~(\ref{eq:sampling}), and estimation of its standard deviation, cf. Eq.~(\ref{eq:std:dev}).
		\item[(B)] {\em Filtering.} Choose the set of nonclassicality filters $\Omega_w(\beta)$, for example the autocorrelation filters in Eqs.~(\ref{eq:ac-ex}) to (\ref{eq:acw}). Multiply the sampled $\Phi(\beta)$ with the filter of width $w$: $\Phi_\Omega(\beta;w) = \Phi(\beta) \Omega_w(\beta)$. The single parameter $w$ is used to optimize the statistical significance of the nonclassical effects to be visualized.
        \item[(C)] {\em Fourier transform.} Calculate the Fourier transform $P_\Omega$ of $\Phi_\Omega$ and its statistical error. If it displays statistically significant negativities, the state is clearly nonclassical. The wider the filter, the more nonclassical effects are visible in the regularized function $P_\Omega$, which is only limited by the increasing sampling noise. 
\end{enumerate}
 
\subsection{Example: Squeezed vacuum state}

For illustration, let us consider a squeezed vacuum state, described by a characteristic function 
\begin{equation}
        \Phi(\beta) = \exp\{-(\beta+\beta^*)^2 V_x/8 + (\beta-\beta^*)^2 V_p/8 + |\beta|^2/2\} ,
\end{equation}
where $V_x$ and $V_p$ are the variances of two orthogonal quadratures and $V_x < 1 < V_p$. The $P$~function is highly singular: It is composed of derivatives up to infinite orders of the Dirac $\delta$~distribution. 
Hence it is extremely difficult to verify the nonclassicality of a squeezed vacuum state in this general sense.

\begin{figure}[h]
        \includegraphics[width=0.40\textwidth]{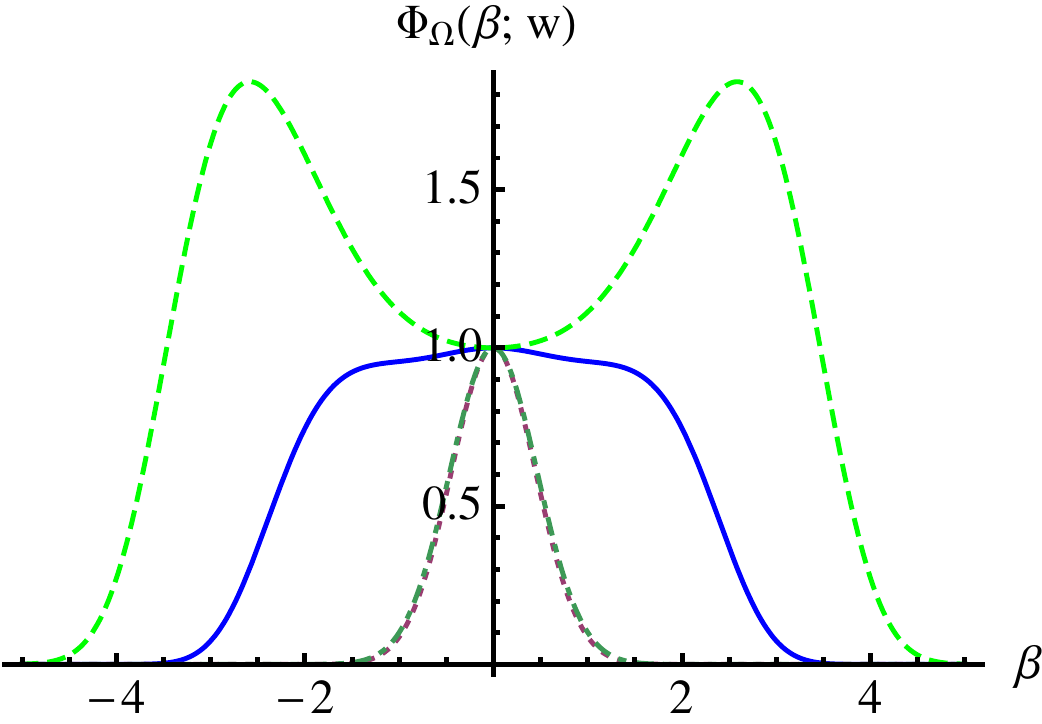} 
        \caption{(Color online) Cross sections of filtered characteristic functions $\Phi_\Omega$ of a squeezed state with $V_x = 0.2$, $V_p = 5.0$. The solid line shows $\Phi_\Omega$ along the squeezed axis with a filter width $w = 1.2$, the dashed line with $w = 1.5$. The narrow curves belong to the unsqueezed axis.}
        \label{fig:CF}
\end{figure}

All $s$-parameterized quasiprobability distributions are either Gaussian or highly singular, and therefore none of them has negativities which can be directly reconstructed. Let us now consider a squeezed state with $V_x = 0.2$ and $V_p = 5.0$, which can be experimentally realized. We construct the filters by Eqs.~(\ref{eq:ac-ex}) to~(\ref{eq:acw}), with a single control parameter $w$. Figure~\ref{fig:CF} shows cross sections of the filtered characteristic functions for two filter widths, $w = 1.2$ and $w=1.5$. The broad and narrow curves correspond to the squeezed and antisqueezed axes, respectively. Without regularization, $\Phi$ grows exponentially in the direction of the squeezed axis, whereas the filtered function $\Phi_\Omega$ is square integrable. 

\begin{figure}[h]
        \includegraphics[width=0.40\textwidth]{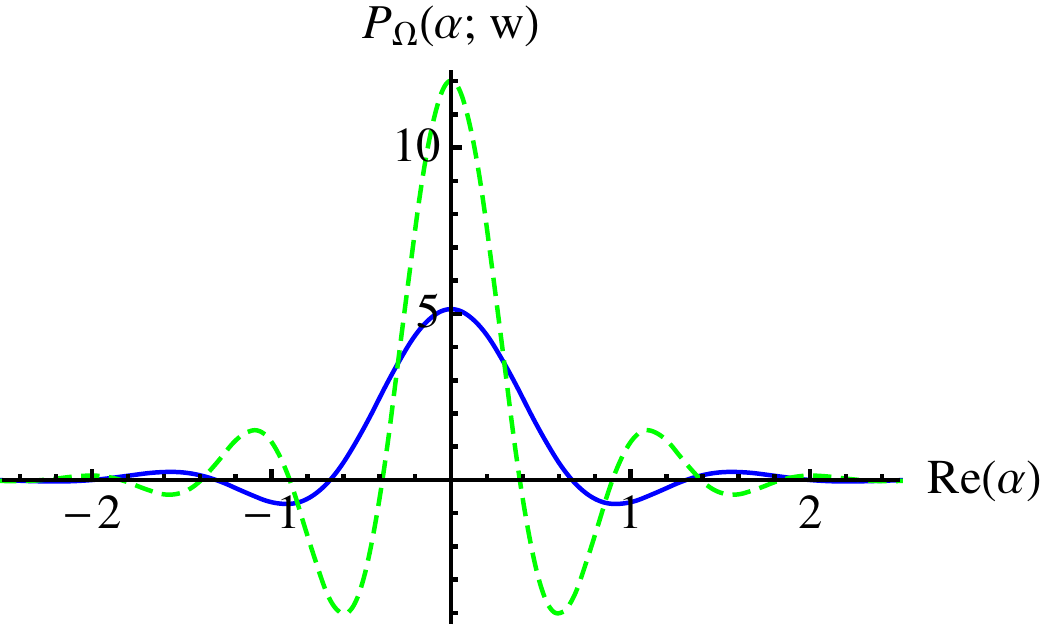} 
        \caption{(Color online) Nonclassicality quasiprobabilities $P_\Omega$ of a squeezed state with $V_x = 0.2$, $V_p = 5.0$ along the squeezed axis. The solid curve is obtained with a filter width $w = 1.2$, the dashed one with $w = 1.5$. The negativities unambiguously visualize the nonclassicality of the squeezed state.} 
        \label{fig:P}
\end{figure}

Cross sections of the resulting nonclassicality quasiprobabilities $P_\Omega$ are given in Fig.~\ref{fig:P}. For both filter widths, they clearly display negativities which have their origin solely in the nonclassicality of the squeezed state. 
The larger the width of the filter, the more pronounced the negativities become. In practice, the filter width is only limited by the experimental uncertainties: It should be sufficiently small to keep the statistical error at a reasonable level.

We stress that the decisive point in our procedure is that the filter has a non-negative Fourier transform. As a consequence, in our approach, the negativities of the filtered function $P_\Omega$ can only be due to the nonclassicality of the quantum state under study. In this respect, our method differs from previous regularizations of the $P$ function, for example, the one by Klauder~\cite{prl-16-534}, where the used filter has negativities in its Fourier transform. Consequently, negativities of the regularized function $P_\Omega$ do not definitely display the nonclassicality of the state. For example, the filtered $P$ function of a coherent state is the displaced Fourier transform of the filter itself. It always shows negativities, even for the only classical pure state. Recently, the $P$~function of a squeezed state has been regularized by such a filter~\cite{science-322-563}. The obtained negativities, however, cannot be interpreted as the nonclassicality of the state itself. 

In another experiment, the (in this case regular) $P$~function of a single-photon added thermal state has been reconstructed by using a rectangular filter~\cite{pra-78-021804}. 
This requires prior knowledge about the state to estimate
the systematic error caused by the regularization, in order to ensure the significance of the nonclassical effects. With the methods introduced here, 
the negativities visualize the nonclassicality without any prior knowledge about the state.

The extension of our methods to several radiation modes is straightforward. The nonclassicality displayed by the nonclassicality quasiprobabilities also includes entanglement. To directly display entanglement, negativities of entanglement quasiprobabilities can be used~\cite{pra-79-042337}. Together with the present method this yields powerful tools for characterizing quantum systems, which are useful for various applications.

\section{Summary}

We have introduced necessary and sufficient conditions for the nonclassicality of a quantum state which can be directly applied in experiments. Universal nonclassicality filters and regular nonclassicality quasiprobabilities have been introduced which display the  nonclassicality of any quantum state without prior knowledge of its properties. We have constructed experimentally useful filter functions which only depend on a single width parameter. The nonclassical properties of a squeezed state have been visualized by negativities of regular functions, which is impossible with $s$-parameterized quasiprobabilities.

\emph{Acknowledgments.} This work was supported by the Deutsche Forschungsgemeinschaft through SFB 652. We are grateful to J. Sperling for valuable comments.

\appendix
\section{General properties of nonclassicality filters} 
\label{app:A}
The approach we are developing shall be applicable to any quantum state, also on the basis of experimental data. This was already demonstrated in the main text. In addition, negativities of regular functions should prove nonclassicality in a one-to-one manner.
\begin{theorem} \label{theorem}
The $P$~function describes a nonclassical state, if and only if the regularized function $P_\Omega(\beta;w)$ shows negativities for a sufficiently large, but finite filter width $w$.
\end{theorem}

\emph{Proof:}
Let us assume that a state given by its characteristic function $\Phi(\beta)$ is nonclassical. Owing to Bochner's theorem, this implies that~\cite{prl-89-283601}:
\begin{eqnarray}
 \exists N \in \mathbb{N}, \beta_1,\ldots,\beta_N \in \mathbb C: \nonumber\\
        D_N = {\rm det}\left\{\left(\Phi(\beta_i-\beta_j)\right)_{i,j=1,\ldots,N}\right\} <  0.
        \label{eq:det:less:zero}
\end{eqnarray}
Let us take a sequence of filters $\Omega_w(\beta)$ which satisfies all the properties~(a)-(c). Then we have for $N$ and $\beta_1,\ldots,\beta_N$ as chosen in Eq.~(\ref{eq:det:less:zero}),
\begin{equation}
  \lim_{w\to\infty} {\rm det}\left\{\left(\Phi(\beta_i-\beta_j)\Omega_w(\beta_i-\beta_j)\right)_{i,j=1,\ldots,N}\right\} = D_N  < 0.
\end{equation}
Since the determinant is a continuous function, there must exist a finite $w_0> 0$ such that 
\begin{equation}
 {\rm det}\left\{(\Phi(\beta_i-\beta_j)\Omega_{w_0}(\beta_i-\beta_j))_{i,j=1,\ldots,N}\right\} < 0.
\end{equation}
Hence, we have found a filter $\Omega_{w_0}(\beta)$ such that the filtered state $\Phi_\Omega(\beta;w_0)$ is nonclassical. Since its Fourier transform $P_\Omega(\beta;w_0)$ is a regular function but not a probability density, it must show negativities to display nonclassicality. 

The other way around, if $\Phi(\beta)$ represents a classical quantum state, its characteristic function satisfies the conditions of Bochner's criterion. Furthermore, each filter $\Omega_w(\beta)$ shall have a non-negative Fourier transform, and hence it also satisfies the conditions of Bochner. Under these assumptions, it is immediately clear that $\Phi_\Omega(\beta;w)$ also satisfies $\Phi_w(-\beta)=\Phi^*_w(\beta)$ and $\Phi_w(0)=1$. Moreover, the matrix $(\Phi_\Omega(\beta_i-\beta_j;w))_{i,j=1\ldots N}$ is the Hadamard product of the matrices $(\Phi(\beta_i-\beta_j))_{i,j=1\ldots N}$ and $(\Omega_w(\beta_i-\beta_j))_{i,j=1\ldots N}$. If the latter two matrices are positive semidefinite, their Hadamard product is also positive semidefinite. Consequently, $\Phi_\Omega(\beta;w)$ satisfies the conditions of Bochner's theorem. Hence, for any classical state with a positive semidefinite $P$~function, the regularized function $P_\Omega(\beta;w)$ is a classical probability distribution showing no negativities. \hfill $\blacksquare$

\section{Decay properties of autocorrelation filters}
\label{app:B}
Here, we prove the following lemma, which has been used for introducing nonclassicality quasiprobabilities.
\begin{lemma} \label{lemma}
Let $u$ be a real positive number and $\omega(\beta)$ a real function which satisfies $C = \| \omega(\beta)e^{u |\beta|^2}\|_2 < \infty$, where $\|\cdot\|_2$ is the $L^2$-norm. Then the autocorrelation function of $\omega(\beta)$,
\begin{equation}
 	\Omega(\alpha)  = \int \omega(\beta)\omega(\alpha+\beta) d^2 \beta,
\end{equation}
satisfies $ \| \Omega(\alpha) e^{v |\alpha|^2}\|_2 < \infty$ for any real $v < \tfrac{u}{2}$.
\end{lemma}

\emph{Proof:} The autocorrelation function can be rewritten in the following way:
\begin{eqnarray}
 	\Omega(\alpha) = \int \omega(\beta)e^{u|\beta|^2} \omega(\alpha+\beta) e^{u|\alpha+\beta|^2} \\ \nonumber \times e^{-u |2\beta + \alpha|^2/2}\, d^2\beta\, e^{-u|\alpha|^2/2}.
\end{eqnarray}
It is bounded from above by
\begin{eqnarray}
 	|\Omega(\alpha)| &\leq& \|\omega(\beta)e^{u|\beta|^2} \omega(\alpha+\beta) e^{u|\alpha+\beta|^2} e^{-u |2\beta + \alpha|^2/2}\|_1 
	\nonumber\\&&\times e^{-u|\alpha|^2/2}
\end{eqnarray}
with $\|\cdot\|_1$ being the $L^1$-norm. Applying H\"older's inequality \cite{Yosida} in the form $\|f g h\|_1 \leq \|f\|_2\|g\|_2\|h\|_\infty$ with $f(\beta)=\omega(\beta)e^{u|\beta|^2}$, $g(\beta)=\omega(\alpha+\beta) e^{u|\alpha+\beta|^2}$ and $\|e^{-u |2\beta + \alpha|^2/2}\|_\infty = 1$ gives
\begin{equation}
 	|\Omega(\alpha)| \leq C^2 e^{-u|\alpha|^2/2}.
\end{equation}
Since $C$ is finite,  we have
\begin{equation}
 \|\Omega(\alpha) e^{v |\alpha|^2}\|_2 \leq  C^2\| e^{(v - u/2) |\alpha|^2}\|_2
\end{equation}
where the right-hand side is finite if $v < \tfrac{u}{2}$. \hfill $\blacksquare$

\emph{Remark.} If $\omega(\beta)$ satisfies the requirements of lemma~\ref{lemma} for all $u > 0$, the same holds also for $\Omega(\beta)$.

\end{document}